\documentclass[pra,twocolumn,amsmath,amssymb,floatfi,showpacs]{revtex4}
\usepackage{graphicx}

\begin{document}
\title{\textbf{Engineering a thermal squeezed reservoir by system energy modulation}}
\author{Ephraim Shahmoon}
\author{Gershon Kurizki}
\affiliation{Department of Chemical Physics, Weizmann Institute of Science, Rehovot, 76100, Israel}
\date{\today}

\begin{abstract}
We show that a thermal reservoir can effectively act as a squeezed reservoir on atoms that are subject to energy-level modulation. For sufficiently fast and strong modulation, for which the rotating-wave-approximation is broken, the resulting squeezing persists at long times. These effects are analyzed by a master equation that is valid beyond the rotating wave approximation. As an example we consider a two-level-atom in a cavity with Lorentzian linewidth, subject to sinusoidal energy modulation. A possible realization of these effects is discussed for Rydberg atoms.
\end{abstract}

\pacs{03.65Yz, 42.50.-p} \maketitle

\section{Introduction}\label{sec1}
A two-level atom (TLA) that interacts with squeezed light fields (or squeezed reservoirs) may exhibit characteristic dynamics and fluorescence properties that depend on the strength and/or phase of the squeezing \cite{DF,SAV}. For example, different decay rates for the two atomic-dipole quadratures \cite{GAR} and subnatural linewidth of resonance fluorescence peaks \cite{CAR}, were predicted. Some of these features persist when the more realistic non-minimum-uncertainty state of thermal squeezed light is considered \cite{VIET}. Recently it has also been shown that the error induced in coherent control of atoms by quantum fluctuations and light-atom entanglement may be either suppressed or enhanced when a squeezed light pulse is used for such control \cite{EFI}. However, a challenging limitation on the observability of most of these phenomena is the rather poor spatial overlap between the dipole radiation pattern of the atom and available paraxial sources of squeezed light. For this reason, atomic interaction with squeezed light in a cavity has been considered \cite{PAR}.

A completely different approach towards observing such phenomena involves emulating the interaction of atoms with squeezed light, while the atom actually interacts with a reservoir in a thermal or vacuum state. This may be achieved by controlling accessible parameters of the  atomic system without affecting the state of the inaccessible reservoir, thereby creating an effectively squeezed reservoir. For example, in \cite{LCZ} two states of a laser-driven four-level-atom are shown to be coupled to an effectively squeezed reservoir whose properties are determined by the laser parameters. More proposals using multilevel atoms can be found in \cite{POL,FL,PAR2}. In \cite{TANAS} it is shown that when a TLA is driven by a strong classical field, while coupled to a thermal reservoir, it may be viewed as coupled to an effectively squeezed reservoir, as long as the Rabi frequency of the driving field is not too small compared to the spectral width of the reservoir.

Here we show that modulations of the level-spacing of a TLA coupled to a thermal reservoir can lead to the same effect as its coupling to a squeezed reservoir. The key principle of our approach, that allows for effective squeezing, is the breakdown of the rotating-wave-approximation (RWA) of the field-atom interaction, resulting in an interaction similar to that obtained for a squeezed reservoir. The RWA implies neglecting the counter-rotating terms in the system-reservoir coupling that oscillate at least as fast as the level-spacing frequency, $\omega_a$, and is valid for times much larger than $\omega_a^{-1}$ \cite{CCT}. However, at shorter times, the counter-rotating terms are responsible for the appearance of terms characteristic of a squeezed reservoir in the master equation of the system, although it is  coupled to a thermal, non-squeezed, reservoir, as was shown in \cite{MAN} for quantum Brownian motion (conversely, when adopting the RWA but considering the interaction with a squeezed field mode, the effect of squeezing is reminiscent of the existence of counter-rotating terms \cite{EFI}). The question is whether modulation of the level-spacing can help preserve the counter-rotating terms even at times much longer than $\omega_a^{-1}$ and thus lead to coupling to an effectively squeezed reservoir. Such modulations have been extensively studied for both short, non-Markovian, time scales and long, Markovian, ones, and shown to drastically modify atomic decay \cite{KUR1}, decoherence \cite{KUR2} and thermodynamics \cite{KUR3}, through either Zeno-like suppression of the coupling to the reservoir or anti-Zeno-like enhancement.

We find that when the level-spacing modulations of a TLA that is coupled to a thermal reservoir, are sufficiently fast and strong to break the RWA, the practically achievable degree of squeezing does not exceed that of a "classically squeezed state" \cite{DF}, namely, a state wherein the effective squeezing parameter of the reservoir, $M$, and its effective mode occupation at $\omega_a$, $N$, satisfy $|M|\leq N$. In such a state the quadrature noise of the reservoir cannot be less than that of the vacuum \cite{DF,MW}. Emulating quantum squeezing, so that $N<|M|\leq\sqrt{N(N+1)}$ \cite{DF}, is possible, in principle, by this method, but rather impractical using present-day technology. In order to preserve the counter-rotating terms at long times, i.e. make them non-oscillatory in time, the required modulation strength and frequency is found to be of the order of $\omega_a$, which makes this scheme suitable for the microwave regime rather than the optical one.

The paper is organized as follows. In Sec. II we provide simple arguments that explain the relation between counter-rotating terms and effective squeezing, and the effect of modulations. Sec. III is devoted to the derivation of a generalized non-Markovian master equation (ME) for a TLA in a thermal reservoir, in the presence of level-spacing modulations and without invoking the RWA. It contains our main general result, presented in Eqs. (\ref{GME},\ref{MN1}), i.e. the emergence of effective squeezing terms in the ME for the system density operator. The general analysis is illustrated in Sec. IV, for the case of an atom coupled to a cavity with a Lorentzian spectrum. In Sec. V we discuss the analytically solvable example of sinusoidal modulation and explicitly derive the corresponding squeezing parameter. Sec. VI includes a discussion of possible observable effects, such as atomic-dipole dephasing and fluorescence, in the context of our modulation scheme. Finally, Sec. VII presents relevant experimental considerations and an example of a possible Rydberg-atom realization of such effects. The conclusions are presented in Sec. VIII.

\section{Why counter-rotating terms induce effective squeezing}\label{sec2}
In this section we wish to explain, in simple terms, our main idea for engineering an effectively squeezed thermal reservoir. We begin by considering the interaction between a TLA with energy $\hbar\omega_a$ and a broadband electromagnetic field with resonant carrier frequency $\omega_a$,
\begin{equation}
H=\hbar d \left(\hat{\sigma}_{+}+\hat{\sigma}_{-}\right)\left(\hat{E}e^{-i\omega_a t}+\hat{E}^{\dag}e^{i\omega_a t}\right),
\end{equation}
where $\hat{E}$ is the positive-frequency envelope of the field, $d$ is the dipole matrix element and $\hat{\sigma}_{\pm}$ are the TLA operators. In the interaction picture we obtain
\begin{equation}
H_I=\hbar d \left(\hat{\sigma}_{+}\hat{E} +\hat{\sigma}_{-}\hat{E}^{\dag}+
\hat{\sigma}_{+}\hat{E}^{\dag}e^{i2\omega_a t}+\hat{\sigma}_{-}\hat{E}e^{-i2\omega_a t}\right),
\end{equation}
where $\hat{E}(t)$ is time-dependent. The last two terms are usually neglected for times $t\gg\omega_a^{-1}$ in the RWA. Without invoking the RWA, this Hamiltonian can be written in the form,
\begin{eqnarray}
H_I&=&\hbar \left(\hat{\sigma}_{+}\hat{F}+\hat{\sigma}_{-}\hat{F}^{\dag}\right)
\nonumber\\
\hat{F}(t)&=&d\left[\hat{E}(t)+\hat{E}^{\dag}(t)e^{i2\omega_a t}\right],
\label{F}
\end{eqnarray}
where $\hat{F}(t)$ may be recognized as a Langevin-like force, due to the reservoir, acting on the atomic-dipole. In the Markov approximation, the Langevin force is assumed delta-correlated in time, hence the normal and anomalous correlations of the force then determine the parameters $M$ and $N$, respectively, as follows:
\begin{eqnarray}
\langle \hat{F}^{\dag}(t)\hat{F}(t+\tau)\rangle&=&\gamma N \delta(\tau),
\nonumber\\
\langle \hat{F}(t)\hat{F}(t+\tau)\rangle&=&\gamma M \delta(\tau),
\label{FF}
\end{eqnarray}
where $\gamma$ is the decay rate of the TLA population due to the reservoir. Let us now compare the origin of $M$ and $N$ in the cases of \emph{(a)} a real squeezed reservoir with RWA, \emph{(b)} thermal reservoir without RWA  and \emph{(c)} the case of modulations.

\emph{Case a: squeezed reservoir + RWA}.--- assume that the broadband field $\hat{E}(t)$ has squeezed field correlations, i.e.
\begin{eqnarray}
\langle \hat{E}(t)\hat{E}^{\dag}(t+\tau)\rangle&=& (n+1) \delta(\tau),
\nonumber\\
\langle \hat{E}^{\dag}(t)\hat{E}(t+\tau)\rangle&=& n \delta(\tau),
\nonumber\\
\langle \hat{E}(t)\hat{E}(t+\tau)\rangle&=& m \delta(\tau).
\label{EE}
\end{eqnarray}
By taking the RWA in Eq. (\ref{F}) we recognize that $\hat{F}=d \hat{E}$, and comparing Eq. (\ref{FF}) to Eq. (\ref{EE}) we obtain $M=d^2 m/\gamma$, $N=d^2 n/\gamma$, i.e. the effective reservoir, $\hat{F}$, is squeezed ($M\neq0$) since the "real" reservoir, $\hat{E}$, is squeezed ($m\neq0$).

\emph{Case b: thermal reservoir without RWA}.--- consider now that the broadband field $\hat{E}(t)$ is thermal, namely it has only normal correlations, $n\neq0$ and $m=0$ in Eq. (\ref{EE}). However, if we assume that the relevant timescales we are interested in satisfy $t\ll\omega_a^{-1}$, then we may set $e^{i2\omega_a t} \approx 1$ in Eq. (\ref{F}) and the effective Langevin force becomes $\hat{F}=d(\hat{E}+\hat{E}^{\dag})$. Plugging this force into Eq. (\ref{FF}) and using Eq. (\ref{EE}) with $m=0$, yields $M=N=d^2 (2n+1)/\gamma$. This shows that \emph{(1)} the system is effectively coupled to a squeezed reservoir due to the existence of the counter-rotating terms, and \emph{(2)} the achievable squeezing by solely breaking the RWA is that of a "classical squeezed state", i.e. the magnitude of the anomalous correlation $M$ does not exceed the normal correlation $N$.

\emph{Case c: thermal reservoir + modulation, for long times}.--- here we take again a thermal reservoir, $n\neq0$ and $m=0$, and take the long-time limit $t\gg \omega_a^{-1}$ only when modulations are considered. The modulated TLA Hamiltonian is taken to be
\begin{equation}
H_A(t)=\frac{1}{2}\hbar \hat{\sigma}_z \left[\omega_a+\delta(t)\right].
\label{HA}
\end{equation}
Then, in the interaction picture $\hat{\sigma}_{-,I}(t)=\hat{\sigma}_{-}e^{-i\omega_a t}\varepsilon^{\ast}(t)$, with
\begin{equation}
\varepsilon(t)=e^{i\int_0^t dt' \delta(t')}=\sum_q \varepsilon_q e^{i\nu_q t},
\label{eps}
\end{equation}
where we assumed that $\varepsilon(t)$ can be written as a Fourier series, and let us take its period to be $2\pi/\omega_a$, i.e. $\nu_q=q\omega_a$ with integer $q$. The Langevin force, Eq. (\ref{F}), becomes,
\begin{equation}
\hat{F}(t)=\sum_q \varepsilon_q d \left[ \hat{E}(t) e^{iq \omega_a t}+\hat{E}^{\dag}(t)e^{i(2+q)\omega_a t}\right].
\label{Fm}
\end{equation}
From the above equation it is evident that in order to obtain squeezing similar to that in case \emph{b}, it is sufficient that the terms  $q=0$ and $q=-2$ exist. Then, at long times $t\gg\omega_a^{-1}$, at which the oscillatory terms $e^{i2\omega_a t}$ vanish, we obtain  $\hat{F}(t)\approx d( \varepsilon_0 \hat{E}+\varepsilon_{-2} \hat{E})$, with $M=\varepsilon_0\varepsilon_{-2}(2n+1)(d^2/\gamma)$ and $N=[|\varepsilon_0|^2 n+|\varepsilon_{-2}|^2(n+1)](d^2/\gamma)$. For $|\varepsilon_0|=|\varepsilon_{-2}|$ we recover case \emph{b} where classical squeezing is achievable. Quantum squeezing however, namely $|M|>N$, is practically more difficult to obtain, though possible in principle, as explained in Appendix A. At this point it also becomes clearer why strong and fast modulation $\delta(t)$ is required: the existence of squeezing is enabled by the $q=-2$ term with frequency $-2\omega_a$. Then $\int dt' \delta(t')$ has to contain a frequency of order $\omega_a$, e.g. $\sin(\omega_a t)$, such that the derivative $\delta(t)$ includes a term with magnitude and frequency $\omega_a$, e.g. $\omega_a \cos(\omega_a t)$.
This summarizes our basic idea of how to create an effective squeezed reservoir by system energy-modulation.

\section{Generalized master equation}\label{sec3}
To provide a more rigorous account of such effects, we derive a master equation (ME) for the TLA density operator, in the presence of electromagnetic field modes in a thermal state, when the TLA level-spacing undergoes modulations. The system+reservoir Hamiltonian reads,
\begin{eqnarray}
H(t)&=&H_A(t)+H_F+H_{AF},
\nonumber \\
H_F&=&\sum_k\hbar \omega_k \hat{a}^{\dag}_k\hat{a}_k,
\nonumber \\
H_{AF}&=&\sum_k\hbar g_k (\hat{\sigma}_{-}+\hat{\sigma}_{+})(\hat{a}^{\dag}_k+\hat{a}_k),
\label{Htot}
\end{eqnarray}
where $k$ are indices of different field modes with corresponding destruction operators $\hat{a}_k$, frequencies $\omega_k$ and dipole couplings $g_k$, and $H_A$ is the one in Eq. (\ref{HA}). In the interaction picture with respect to $H_A+H_F$, the Hamiltonian becomes
\begin{eqnarray}
H_I(t)&=&\hbar \left[\hat{\sigma}_{-}\hat{F}_a^{\dag}(t) \varepsilon^{\ast}(t) +\hat{\sigma}_{-}\tilde{F}_a\varepsilon^{\ast}(t) +  \mathrm{h.c.}\right],
\nonumber\\
\hat{F}_a(t)&=&\sum_k g_k \hat{a}_k e^{-i(\omega_k-\omega_a)},
\nonumber\\
\tilde{F}_a(t)&=&\sum_k g_k \hat{a}_k e^{-i(\omega_k+\omega_a)},
\label{Fa}
\end{eqnarray}
with $\varepsilon(t)$ from Eq. (\ref{eps}). Note that the tilded operators $\tilde{F}_a(t),\tilde{F}^{\dag}_a(t)$ are the ones that are neglected in the RWA without modulation. In the Born approximation, namely, when neglecting the system-reservoir correlations, the ME for the TLA density operator $\rho$ is \cite{CARb},
\begin{equation}
\dot{\rho}(t)=-\frac{1}{\hbar^2}\int_0^t dt' \mathrm{tr}_F \left([H_I(t),[H_I(t'),\rho(t')\rho_F]]\right),
\label{ME1}
\end{equation}
where $\rho_F$ is the stationary density operator of the reservoir (the field). The double commutator inside the trace over field degrees of freedom contains four terms,
\begin{eqnarray}
&&H_I(t)H_I(t')\rho(t')\rho_F - H_I(t)\rho(t')\rho_F H_I(t')
\nonumber\\
&&-H_I(t')\rho(t')\rho_F H_I(t)+\rho(t')\rho_F H_I(t')H_I(t),
\label{ABCD}
\end{eqnarray}
each of which contains 16 terms. In order to illustrate the main points of the derivation, let us focus on the second term in (\ref{ABCD}), and, more specifically, on three representative terms out of its 16 terms
\begin{eqnarray}
&&\hat{\sigma}_{+}\rho(t')\hat{\sigma}_{-}\hat{F}_a(t)\rho_F\hat{F}_a^{\dag}(t')\varepsilon(t)\varepsilon^{\ast}(t'),
\nonumber\\
&&\hat{\sigma}_{+}\rho(t')\hat{\sigma}_{+}\hat{F}_a(t)\rho_F\hat{F}_a(t')\varepsilon(t)\varepsilon(t'),
\nonumber\\
&&\hat{\sigma}_{+}\rho(t')\hat{\sigma}_{+}\tilde{F}_a^{\dag}(t)\rho_F\hat{F}_a(t')\varepsilon(t)\varepsilon(t').
\label{B}
\end{eqnarray}
The first two terms do not contain tilded operators, so they exist also in the RWA without modulations. When the trace over the field is taken, it can be seen that the first term includes the normal correlator $\langle \hat{F}_a^{\dag}(t')\hat{F}_a(t)\rangle=\mathrm{tr}_F(\hat{F}_a^{\dag}(t')\hat{F}_a(t)\rho_F)$, which exists for a thermal reservoir. This term contributes to the $\hat{\sigma}_{+}\rho\hat{\sigma}_{-}$ term in the ME, typical of a thermal reservoir \cite{MW,CARb}. The second term involves the anomalous correlator  $\langle \hat{F}_a(t')\hat{F}_a(t)\rangle$ which exists only for a squeezed reservoir and contributes to the squeezing term $\hat{\sigma}_{+}\rho\hat{\sigma}_{+}$ in the ME \cite{MW}. However, in our case we assume a thermal, non-squeezed, reservoir, and this correlator vanishes, so that indeed, for RWA terms, no squeezing is expected when a thermal reservoir is considered. The third term in (\ref{B}) is the most interesting for our purposes. It does not exist in the RWA as it contains a tilded operator, yet, if the modulations preserve it at long times, it contributes to the squeezing terms of the ME, $\hat{\sigma}_{+}\rho\hat{\sigma}_{+}$. We calculate its associated correlator $\langle \hat{F}_a(t')\tilde{F}_a^{\dag}(t)\rangle$ using Eq. (\ref{Fa}),
\begin{eqnarray}
&&\langle \hat{F}_a(t')\tilde{F}_a^{\dag}(t)\rangle=\sum_k\sum_{k'}g_k g_{k'}^{\ast}e^{-i(\omega_k-\omega_a)t'}e^{i(\omega_k'+\omega_a)t}\langle \hat{a}_k \hat{a}_{k'}^{\dag}\rangle
\nonumber\\
&&=e^{i2\omega_a t}\sum_k|g_k|^2e^{-i(\omega_k+\omega_a)(t-t')}(n_k+1)
\nonumber\\
&&= e^{i2\omega_a t}\int d\omega D(\omega)|g(\omega)|^2e^{-i(\omega+\omega_a)(t-t')}[n(\omega)+1].
\label{B30}
\end{eqnarray}
Here we used $\langle \hat{a}_k^{\dag}\hat{a}_{k'}\rangle=\delta_{kk'}n_k$ for a thermal state of the field, where $n_k$ is the Planck distribution for frequency $\omega_k$, and we defined the density of field modes by $\sum_k\rightarrow\int d\omega D(\omega)$. We further define the coupling spectrum of the reservoir,
\begin{equation}
G_0(\omega)=D(\omega)|g(\omega)|^2.
\label{G0}
\end{equation}
Reexamining the expressions in (\ref{B}) and the ME (\ref{ME1}), we need to multiply this correlator by the modulation functions and integrate over time,
\begin{eqnarray}
&&e^{i2\omega_a t}\varepsilon(t)\int_0^t dt'\varepsilon(t')\int d\omega G_0(\omega)[1+n(\omega)]e^{-i(\omega+\omega_a)(t-t')}
\nonumber\\
&&=\sum_q\sum_{q'} \varepsilon_q \varepsilon_{q'}e^{i(\nu_q+\nu_{q'}+2\omega_a)t} \times
\nonumber\\
&&\int d\omega G_0(\omega)[1+n(\omega)] \int_0^t dt' e^{-i(\omega+\omega_a+\nu_q)(t-t')},
\label{B3a}
\end{eqnarray}
where the definition of $\varepsilon(t)$ from (\ref{eps}) was used. Here $\rho(t')$ was taken out of the integral by assuming that $t$ is much shorter than the typical timescale for changes in the system, i.e. $\rho(t')\approx \rho(t)$. In cases considered later, when the Markov approximation is taken and the resulting ME has time-independent coefficients, we may view this short $t$ as a coarse-graining time and the validity of the ME is then extended to long times \cite{CCT}. In Appendix B we show that the term in (\ref{B3a}) can be written as
\begin{eqnarray}
&&\sum_q\sum_{q'} \varepsilon_q \varepsilon_{q'}e^{i(\nu_q+\nu_{q'}+2\omega_a)t} \times
\nonumber\\
&&\int d\omega \delta_t(\omega) \left[\frac{1}{2}\gamma_n(\omega-\omega_a-\nu_q)-i\Delta_n(\omega-\omega_a-\nu_q )\right],
\nonumber\\
\label{B3b}
\end{eqnarray}
where $\delta_t(\omega)=t \frac{\sin(\omega t)}{\omega t}$ is a sinc function that approaches a delta function as $t$ gets larger, and
\begin{eqnarray}
\gamma_n(\omega)&=&2\pi G_0(\omega)[1+n(\omega)],
\nonumber\\
\Delta_n(\omega)&=&P\int_{-\infty}^{\infty} d\omega'\frac{ G_0(\omega')[1+n(\omega')]}{\omega'-\omega},
\label{B3c}
\end{eqnarray}
with $P$ denoting the principal value. Repeating the procedure shown in Eqs. (\ref{B30},\ref{B3a},\ref{B3b}) for all terms of Eqs. (\ref{ME1},\ref{ABCD}), we find the following generalized non-Markovian ME,
\begin{eqnarray}
\dot{\rho}&=&-i[\frac{1}{2}\Delta\hat{\sigma}_z,\rho]+\frac{\gamma}{2}N[2\hat{\sigma}_{+}\rho\hat{\sigma}_{-}-
\rho\hat{\sigma}_{-}\hat{\sigma}_{+}-\hat{\sigma}_{-}\hat{\sigma}_{+}\rho]
\nonumber\\
&+& \frac{\gamma}{2}(N+1)[2\hat{\sigma}_{-}\rho\hat{\sigma}_{+}-
\rho\hat{\sigma}_{+}\hat{\sigma}_{-}-\hat{\sigma}_{+}\hat{\sigma}_{-}\rho]
\nonumber\\
&-&\gamma M \hat{\sigma}_{+}\rho\hat{\sigma}_{+}-\gamma M^{\ast}\hat{\sigma}_{-}\rho\hat{\sigma}_{-}.
\label{GME}
\end{eqnarray}
The first term of the ME is a Hamiltonian term containing the TLA energy correction $\hbar \Delta$, and would not be of interest in the following. The second and third terms describe absorption and emission from and to the photon field, respectively, typical of damping by a thermal reservoir, where $\gamma$ is the decay rate and $N$ the reservoir population.

The last two terms, which are the most interesting for our discussion, appear to be squeezed-reservoir damping terms. Although we assumed a thermal reservoir, they exist due to the modulation. The ME coefficients, $\Delta,\gamma,N$ are real (whereas $M$ is complex) and are generally time-dependent and affected by modulation:
\begin{widetext}
\begin{eqnarray}
\Delta&=&\sum_{qq'}\frac{1}{2}\left[\varepsilon_q\varepsilon_{q'}^{\ast}e^{i(\nu_q-\nu_{q'})t}+\varepsilon_q^{\ast}\varepsilon_{q'}e^{-i(\nu_q-\nu_{q'})t} \right] \int_{-\infty}^{\infty}d\omega\delta_t(\omega)\left[\Delta_T(\omega+\omega_a+\nu_q)-\Delta_T(\omega-\omega_a-\nu_q)\right]
\nonumber\\
&&-i\sum_{qq'}\frac{1}{2}\left[\varepsilon_q\varepsilon_{q'}^{\ast}e^{i(\nu_q-\nu_{q'})t}-\varepsilon_q^{\ast}\varepsilon_{q'}e^{-i(\nu_q-\nu_{q'})t} \right] \int_{-\infty}^{\infty}d\omega\delta_t(\omega)\frac{1}{2}\left[\gamma_T(\omega+\omega_a+\nu_q)+\gamma_T(\omega-\omega_a-\nu_q)\right],
\nonumber\\
\gamma N&=&\sum_{qq'}\left[\varepsilon_q\varepsilon_{q'}^{\ast}e^{i(\nu_q-\nu_{q'})t}+\varepsilon_q^{\ast}\varepsilon_{q'}e^{-i(\nu_q-\nu_{q'})t} \right]
\int_{-\infty}^{\infty}d\omega\delta_t(\omega)\frac{1}{2}\gamma_T(\omega-\omega_a-\nu_q)
\nonumber \\
&& -i\left[\varepsilon_q\varepsilon_{q'}^{\ast}e^{i(\nu_q-\nu_{q'})t}-\varepsilon_q^{\ast}\varepsilon_{q'}e^{-i(\nu_q-\nu_{q'})t} \right]
\int_{-\infty}^{\infty}d\omega\delta_t(\omega)\Delta_T(\omega-\omega_a-\nu_q),
\nonumber\\
\gamma
(N+1)&=&\sum_{qq'}\left[\varepsilon_q\varepsilon_{q'}^{\ast}e^{i(\nu_q-\nu_{q'})t}+\varepsilon_q^{\ast}\varepsilon_{q'}e^{-i(\nu_q-\nu_{q'})t} \right]
\int_{-\infty}^{\infty}d\omega\delta_t(\omega)\frac{1}{2}\gamma_T(\omega+\omega_a+\nu_q)
\nonumber\\
&&+i\left[\varepsilon_q\varepsilon_{q'}^{\ast}e^{i(\nu_q-\nu_{q'})t}-\varepsilon_q^{\ast}\varepsilon_{q'}e^{-i(\nu_q-\nu_{q'})t} \right]
\int_{-\infty}^{\infty}d\omega\delta_t(\omega)\Delta_T(\omega+\omega_a+\nu_q),
\nonumber\\
\gamma M&=&\sum_{qq'} \varepsilon_q \varepsilon_{q'}e^{i(\nu_q+\nu_{q'}+2\omega_a)t}
\int_{-\infty}^{\infty} d\omega \delta_t(\omega) [\frac{1}{2}\gamma_T(\omega+\omega_a+\nu_q)+\frac{1}{2}\gamma_T(\omega-\omega_a-\nu_q)
\nonumber \\
&&+i\Delta_T(\omega+\omega_a+\nu_q )-i\Delta_T(\omega-\omega_a-\nu_q)],
\nonumber\\
\label{MN1}
\end{eqnarray}
\end{widetext}
where we used the temperature-dependent response of the reservoir
\begin{eqnarray}
\gamma_T(\omega)&=&2\pi G_T(\omega),
\nonumber\\
\Delta_T(\omega)&=&P\int_{-\infty}^{\infty} d\omega'\frac{ G_T(\omega')}{\omega'-\omega},
\nonumber\\
G_T(\omega)&=&G_0(\omega)[1+n(\omega)]+G_0(-\omega)n(-\omega).
\label{GT}
\end{eqnarray}
The effect of the modulation is clear from Eq. (\ref{MN1}). At long times, when $\delta_t(\omega)$ is treated as Dirac delta, the modulation enables a scan through the frequency response of the reservoir $\gamma_T(\omega)+i\Delta_T(\omega)$ and thus change the effective coupling to it. This was also concluded in \cite{KUR2}, where modulations were shown to be useful for decoherence control. As for the squeezing, $\gamma M$, let us first consider it without modulation, i.e. when $\varepsilon_0=1,\nu_0=0, \varepsilon_{q\neq0}=0$. Then $\gamma M$ is fast-oscillating and negligible for $t\gg \omega_a^{-1}$, as it should in the RWA. However, by choosing the right modulation frequencies $\nu_q$ this term may become important at long times as well, as was explained in Sec. II above.

The above ME may be viewed as a generalization of previous treatments \cite{KUR1,KUR2}: \emph{(1)} here the principal-value terms $\Delta_T(\omega_a+\nu_q)$ are not neglected, and can in fact become important for reservoir spectra $G_T(\omega)$ which are asymmetric around $\omega_a+\nu_q$. \emph{(2)} The ME is written in operator from, rather than as a Bloch-equation for the TLA density matrix elements, which makes it easy to generalize to other systems, e.g. an harmonic oscillator (see Sec. VIII).

\section{Lorentzian reservoir: atom in a cavity}\label{sec4}
Up to this point we have not specified what are the field modes which make up the reservoir. Let us now consider an atom in a resonant cavity in the "bad-cavity" limit. The interaction of the atom with a cavity-mode with envelope $\tilde{a}$ can be written in the interaction picture as
\begin{equation}
H_I=\hbar g[\tilde{a}(t)e^{-i\omega_a t}+\tilde{a}^{\dag}(t)e^{i\omega_a t}](\hat{\sigma}_{+}e^{i\omega_a t}+\hat{\sigma}_{-}e^{-i\omega_a t}),
\label{JCM}
\end{equation}
with dipolar coupling $g$, so that the Langevin force acting on the atom in analogy with Eq. (\ref{Fa}) is $\hat{F}_a=g \tilde{a}(t)$. The cavity-mode is damped by the coupling to outside modes and its correlation functions decay exponentially (see Appendix C for details),
\begin{eqnarray}
\langle \tilde{a}(t)\tilde{a}^{\dag}(t+\tau)\rangle&=&(n_a+1)e^{-\frac{\kappa}{2} |\tau|}
\nonumber\\
\langle \tilde{a}^{\dag}(t)\tilde{a}(t+\tau)\rangle&=&n_a e^{-\frac{\kappa}{2} |\tau|},
\label{a}
\end{eqnarray}
where $n_a=n(\omega_a)$ and $\kappa$ is the width of the cavity-mode due to the coupling to outside modes, assumed here to satisfy $\kappa\gg g$ (bad-cavity limit) and $\kappa\ll\omega_a$ (see Appendix C).
\subsection{Lorentzian response}
From the Fourier transform of the correlation functions above we can deduce the Lorentzian spectrum of the cavity-mode, which makes up the reservoir for the atom,
\begin{eqnarray}
G_0(\omega)[n(\omega)+1]&\rightarrow& g^2 (n_a+1) \frac{\kappa}{(\frac{\kappa}{2})^2+(\omega-\omega_a)^2},
\nonumber\\
G_0(\omega)n(\omega)&\rightarrow& g^2 n_a \frac{\kappa}{(\frac{\kappa}{2})^2+(\omega-\omega_a)^2}.
\label{Ga}
\end{eqnarray}
Using this spectrum in Eq. (\ref{GT}) we find the temperature-dependent response of the reservoir
\begin{eqnarray}
\gamma_T(\omega)&=& \gamma_c \left[(n_a+1) \frac{1}{1+4(\frac{\omega-\omega_a}{\kappa})^2}+ n_a \frac{1}{1+4(\frac{\omega+\omega_a}{\kappa})^2}\right],
\nonumber\\
\Delta_T(\omega)&=&\gamma_c \left[(n_a+1)\frac{\kappa^2(\omega_a-\omega)}{4(\omega_a-\omega)^2\kappa+\kappa^2}\right.
\nonumber\\
&&\left.-n_a\frac{\kappa^2(\omega_a+\omega)}{4(\omega_a+\omega)^2\kappa+\kappa^2}\right],
\nonumber\\
\gamma_c&\equiv&2\pi G_0(\omega_a)=2\pi\frac{4g^2}{\kappa},
\label{GTa}
\end{eqnarray}
where $\gamma_c$ is the TLA damping rate to the cavity reservoir without modulation.
The real and imaginary response functions obey the Kramers-Kronig relation and are plotted in Fig. 1.
\begin{figure}
\begin{center}
\includegraphics[scale=0.20]{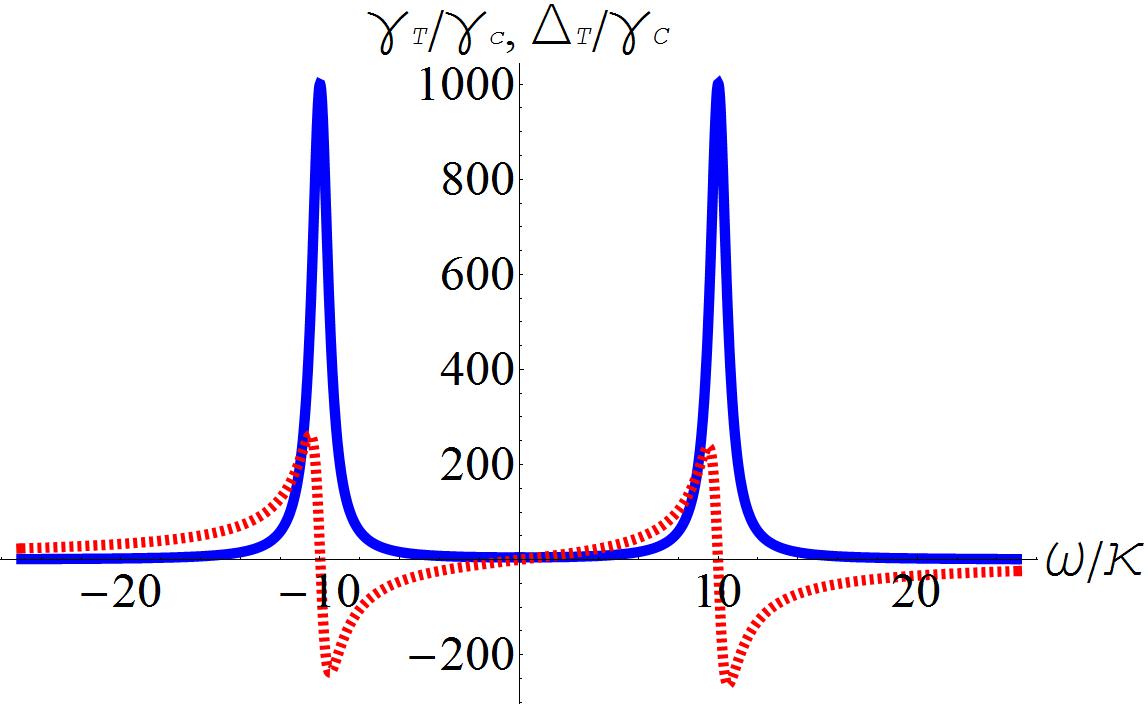}
\caption{\small{(color online). Temperature dependent response of the cavity reservoir, $\gamma_T(\omega)$ (blue solid line) and $\Delta_T(\omega)$ (red dashed line), Eq. (\ref{GTa}). Here $\gamma_c$ is the atomic decay rate to the cavity reservoir in the absence of modulation, and $\omega_a=10\kappa$ is taken.
 }} \label{fig1}
\end{center}
\end{figure}

\subsection{Long-time approximation}
Going back to Eq. (\ref{MN1}) for the ME coefficients, we recall that the sinc function $\delta_t(\omega)$ is of width $1/t$. The typical scale of variations of the  response functions (\ref{GTa}) is $\kappa$, so by assuming $t\gg\kappa^{-1}$ we may take $\delta_t(\omega)\approx \delta(\omega)$ in the integrals over the response functions. This is the well-known Markov approximation, which means that we coarse-grain over timescales of the order $\kappa^{-1}$, such that all the observed phenomena should be slower. Moreover, we are also interested in timescales much longer than $\omega_a^{-1}$, as in the RWA. In Sec. II we saw that in order to obtain effective reservoir squeezing for $t\gg\omega_a^{-1}$, the modulation frequencies $\nu_q$ should be at least of the order $\omega_a$. This means that we should keep only the non-oscillatory terms in the ME coefficients, i.e. $\nu_q=\nu_{q'}$ in $\gamma N, \gamma (N+1)$ and $\nu_q+\nu_{q'}+2\omega_a=0$ in $\gamma M$. To sum up, taking the approximation $t\gg\kappa^{-1},\omega_a^{-1}$, we obtain for these ME coefficients,
\begin{eqnarray}
\gamma N&\approx& \sum_{q}|\varepsilon_q|^2 \gamma_T(-\omega_a-\nu_q),
\nonumber\\
\gamma (N+1)&\approx& \sum_{q}|\varepsilon_q|^2 \gamma_T(\omega_a+\nu_q),
\nonumber\\
\gamma M&\approx&\tilde{\sum}_{qq'}\varepsilon_q \varepsilon_{q'}e^{i(\nu_q+\nu_{q'}+2\omega_a)t} \times
\nonumber\\
&& [\frac{1}{2}\gamma_T(\omega_a+\nu_q)+\frac{1}{2}\gamma_T(-\omega_a-\nu_q)
\nonumber\\
&& +i\Delta_T(\omega_a+\nu_q)-i\Delta_T(-\omega_a-\nu_q )],
\nonumber\\
\label{MN2}
\end{eqnarray}
where $\tilde{\sum}_{qq'}$ denotes the sum with the constraint $\nu_q+\nu_{q'}+2\omega_a=0$. Here the above coefficients are time-independent and the ME, Eq. (\ref{GME}), becomes a Markovian ME of a system damped by an effective thermal squeezed reservoir, which is valid at long times.

\section{Sinusoidal modulation}\label{sec5}
In order to illustrate our scheme for engineering an effective squeezed reservoir for the TLA, let us take the analytically solvable example of a sinusoidal modulation of the TLA level spacing,
\begin{equation}
\delta(t)=z m\omega_a[1-\sin(m\omega_a t)],
\label{delta}
\end{equation}
where $z$ and $m$ are positive parameters. Introducing the above modulation in Eq. (\ref{eps}), and using the identity $e^{i z\cos\phi}=\sum_{q=-\infty}^{\infty} i^q J_q(z) e^{i q \phi}$ with integer $q$, we obtain
\begin{eqnarray}
\varepsilon_q&=&e^{-i z} i^q J_q(z),
\nonumber\\
\nu_q&=&(z+q)m\omega_a.
\label{epssin}
\end{eqnarray}
The constraint $\nu_q+\nu_{q'}+2\omega_a=0$ imposed on the $\gamma M$ term in the long-time approximation now becomes $q'=-(2/m)-2z-q$, and since $q',q$ are integers, we get the following constraints on the parameters $z$ and $m$,
\begin{equation}
z=\frac{\mathrm{integer}}{2} \quad ;  \quad m=\frac{2}{\mathrm{integer}}.
\label{zm}
\end{equation}
The expressions for the coefficients from Eq. (\ref{MN2}) now become,
\begin{eqnarray}
&&\gamma N\approx \sum_{q=-\infty}^{\infty}J_q^2(z) \gamma_T(-\omega_a[1+m z+m q]),
\nonumber\\
&&\gamma (N+1)\approx \sum_{q=-\infty}^{\infty}J_q^2(z) \gamma_T(\omega_a[1+m z+m q]),
\nonumber\\
&&\gamma M\approx\sum_{q=-\infty}^{\infty}e^{-i2z}(-1)^{\frac{1}{m}+z} J_q(z)J_{-q-2z-\frac{2}{m}}(z)\times
\nonumber\\
&& [\frac{1}{2}\gamma_T(\omega_a[1+m z+m q])+\frac{1}{2}\gamma_T(-\omega_a[1+m z+m q])
\nonumber\\
&&+ i\Delta_T(\omega_a[1+m z+m q])-i\Delta_T(-\omega_a[1+m z+m q])].
\nonumber\\
\label{MNs}
\end{eqnarray}
Considering the Lorentzian reservoir from (\ref{GTa}) with $\omega_a=10\kappa$ and $n_a=10^3$, let us choose for example $z=1$ and $m=2$. It is enough to take $q$ from $-20$ to $20$ in order for the sums in Eq. (\ref{MNs}) to converge, and we obtain, $\gamma N=207.49 \gamma_c$, $\gamma (N+1)=207.67 \gamma_c$, $\gamma M=(92.08 - i42.14 )\gamma_c$ and $|\gamma M|=101.26\gamma_c$. This demonstrates an appreciable effect due to modulations, as $|M|$ becomes about half of $N$. On the other hand, we also verified that for $z=0$ we return to the RWA results without modulation, namely, $\gamma N=\gamma_c n_a,\gamma (N+1)=\gamma_c (n_a+1)$ and $\gamma M=0$. In Fig. 2  we plot, with the same $\omega_a$, $n_a$ and as a function of the parameters $z$ and $m$, the difference between $\gamma|M|$ and $\gamma N$. This quantity determines the increase of the atomic-dipole dephasing rate and broadening of the fluorescence line shape, as discussed in Sec. VI below. As this difference becomes smaller, the increase and broadening become smaller compared to the non-squeezed reservoir case. Figs. 2a and 2b present this quantity as a function of $z$ for fixed $m$ values, $m=2$ and $m=1$ respectively, whereas in Fig. 2c it is plotted as a function of $m$ for $z=1$. It appears that $m=2$ is optimal, and in the range of $z$ values considered, $z=9.5$ provides the smallest result, $\gamma N-\gamma|M|=0.128\gamma_c$ with $\gamma N=0.683\gamma_c$. This suggests that indeed $|M|$ does not exceed $N$, i.e. we get "classical squeezing" by breaking the RWA, as argued in Sec. II.
\begin{figure}
\begin{center}
\includegraphics[scale=0.18]{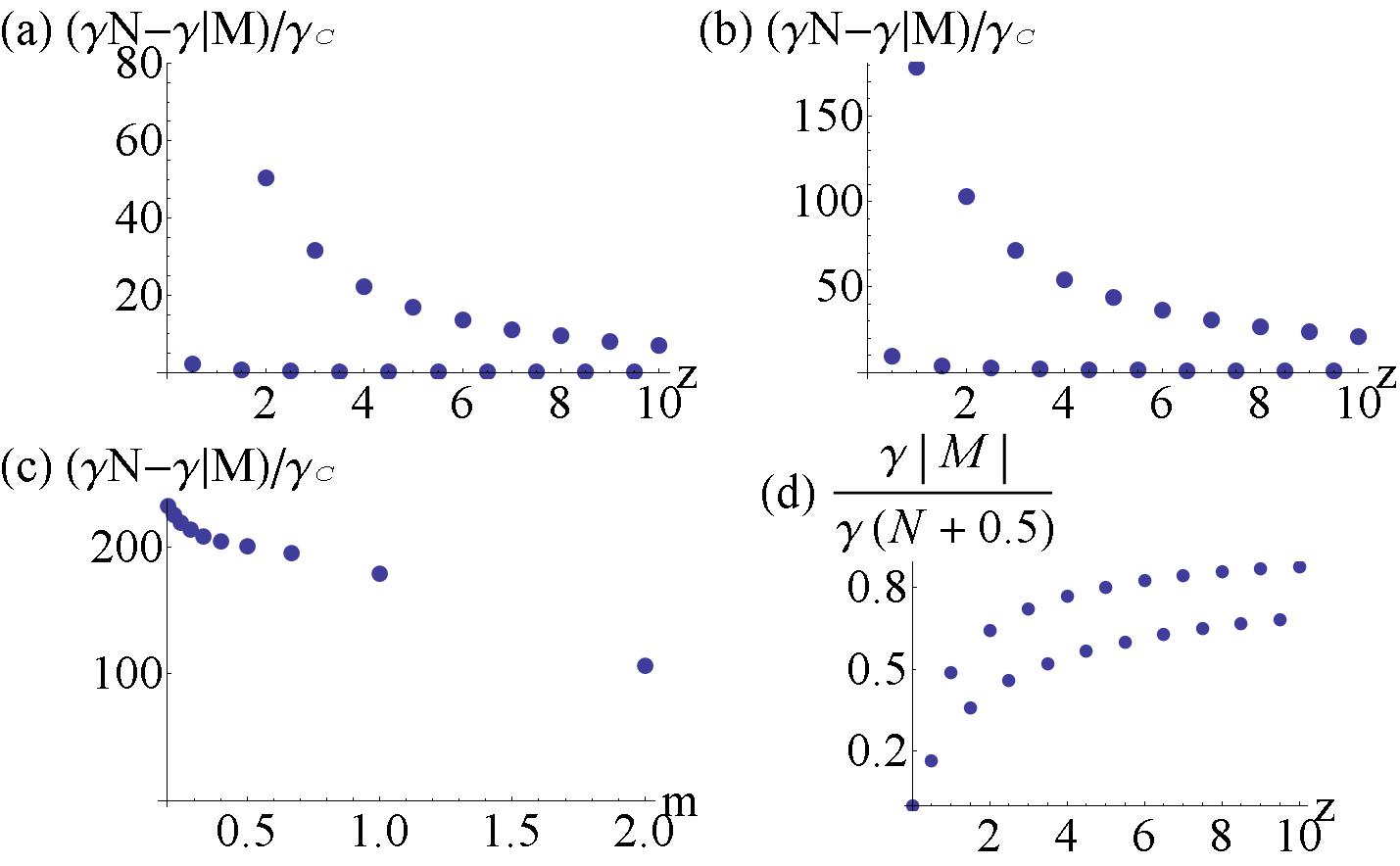}
\caption{\small{(color online). Effective reservoir squeezing $\gamma|M|$ as a function of modulation parameters $z$ and $m$, Eqs. (\ref{delta},\ref{zm}). Taking $\omega_a=10\kappa$ and $n_a=10^3$ in Eqs. (\ref{GTa},\ref{MNs}), the difference between $N$ and $M$ is plotted in (a), (b) and (c) and apparently cannot go lower than the "classical squeezing" bound $|M|\leq N$, as argued in Sec. II. (a) m=2, (b) m=1, (c) z=1. (d) The ratio $|M|/(N+0.5)$, that determines the difference between the two atomic-dipole quadrature dephasing rates, is plotted for $m=2$.
 }} \label{fig2}
\end{center}
\end{figure}

\section{Possible observable effects}\label{sec6}
In this section we wish to describe possible observable manifestations of the effect of modulations and effective squeezing. These include two distinct dephasing rates of the atomic-dipole quadratures, fluorescence and resonance fluorescence.

\subsection{Atomic dipole dephasing}
From the master equation (ME), Eq. (\ref{GME}), we obtain the equations of motion for expectation values of the atomic operators,
\begin{eqnarray}
\langle \dot{\hat{\sigma}}_{+}\rangle&=&-\gamma\left(N+\frac{1}{2}\right)\langle\hat{\sigma}_{+}\rangle-\gamma M^{\ast}\langle \hat{\sigma}_{-}\rangle,
\nonumber\\
\langle \dot{\hat{\sigma}}_{z}\rangle&=&-\gamma\left(2N+1\right)\langle\hat{\sigma}_{z}\rangle-\gamma,
\label{sigp}
\end{eqnarray}
and $\langle \dot{\hat{\sigma}}_{-}\rangle=\langle \dot{\hat{\sigma}}_{+}\rangle^{\ast}$. Denoting $M=|M|e^{i2\varphi}$, we use the squeezing phase $\varphi$ to define the atomic-dipole quadratures as $\hat{\sigma}_x=e^{i\varphi}\hat{\sigma}_{+}+e^{-i\varphi}\hat{\sigma}_{-}$ and $\hat{\sigma}_y=-ie^{i\varphi}\hat{\sigma}_{+}+ie^{-i\varphi}\hat{\sigma}_{-}$, and obtain
\begin{eqnarray}
\langle \dot{\hat{\sigma}}_{x}\rangle&=&-\gamma\left(N-|M|+\frac{1}{2}\right)\langle\hat{\sigma}_{x}\rangle\equiv-\gamma_x\langle\hat{\sigma}_{x}\rangle,
\nonumber\\
\langle \dot{\hat{\sigma}}_{y}\rangle&=&-\gamma\left(N+|M|+\frac{1}{2}\right)\langle\hat{\sigma}_{y}\rangle\equiv-\gamma_y\langle\hat{\sigma}_{y}\rangle.
\label{sigx}
\end{eqnarray}
The above result, shown previously by Gardiner \cite{GAR}, reveals that the two different quadratures may decay with very different rates when the reservoir is squeezed, $|M|\neq0$. In our case, $|M|$ does not exceed $N$, so that the slower rate $\gamma_x$ is bounded by the vacuum rate $\gamma/2$. Note that the effect of modulation on the quadratures' dynamics is twofold: not only does it induce squeezing and hence two distinct decay rates, but it may also change $N$ and $\gamma$ from their thermal-reservoir values without modulation, $n_a$ and $\gamma_c$ respectively. In cases where the dynamics can be measured, they may reveal both effects in the most direct way. Fig. 2d portraits the ratio $\frac{\gamma|M||}{\gamma(N+0.5)}$ for $m=2$ and as a function of $z$. As can be seen in Eq. (\ref{sigx}), this ratio determines the difference between the two decay rates. For instance, when $z=1$ we obtain $\gamma_x=106\gamma_c$, $\gamma_y=309\gamma_c$, whereas for $z=3$ we get $\gamma_x=32$, $\gamma_y=195$, and the corresponding ratios for these $z=1,3$ cases are $\frac{\gamma|M||}{\gamma(N+0.5)}=0.49,0.72$.

\subsection{Atom fluorescence}
Making use of Eq. (\ref{sigx}) and the quantum regression theorem \cite{CARb}, it is easy to obtain the correlation function $C(t)$ and the corresponding spectrum $S(\omega)=\frac{1}{\pi}\int_{-\infty}^{\infty}dte^{-i\omega t}C(t)$ for the atom in steady state,
\begin{eqnarray}
&&C(t)= \langle\hat{\sigma}_{+}(0)\hat{\sigma}_{-}(t)\rangle=\frac{1}{2}\frac{N}{2N+1}\left[e^{-\gamma_x|t|}+e^{-\gamma_y|t|}\right],
\nonumber\\
&&S(\omega)=\frac{1}{\pi}\frac{N}{2N+1}\left[ \frac{\gamma_x}{\gamma_x^2+\omega^2}+\frac{\gamma_y}{\gamma_y^2+\omega^2} \right],
\label{corr}
\end{eqnarray}
where $\omega$ here is shifted by $\omega_a$, i.e. the spectrum is actually centered around $\omega_a$. This spectrum is proportional to that of the fluorescent light emitted by the atom and it reveals both effects of the modulation. First of all, even if squeezing was absent and the spectrum would still be a Lorentzian of width $\gamma (N+0.5)$, the modulation changes it from  the unmodulated thermal value, $(n_a+0.5)\gamma_c=1000.5\gamma_c$. Perhaps the more interesting effect however, is that of squeezing \cite{GAR}, namely the spectrum is actually not a single Lorentzian, rather it is a sum of two Lorentzians with different widths $\gamma_x$ and $\gamma_y$. As in the dipole dephasing case, the effect becomes more apparent as the two widths become more distinct. In Fig. 3  we present, for the case $m=2,z=3$, the comparison of the fluorescence spectrum with and without modulation. We also show the non-Lorentzian shape of the spectrum by comparing it to a Lorentzian of width $\gamma (N+0.5)$ in the same case.
\begin{figure}
\begin{center}
\includegraphics[scale=0.25]{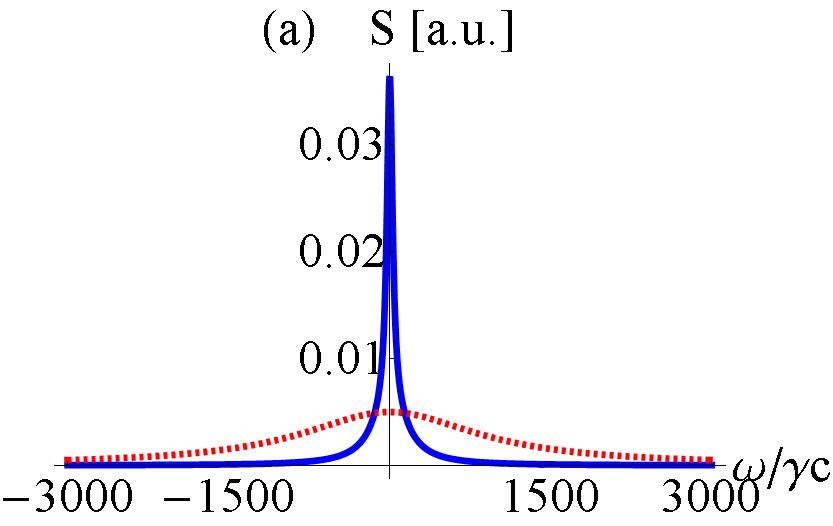}
\includegraphics[scale=0.25]{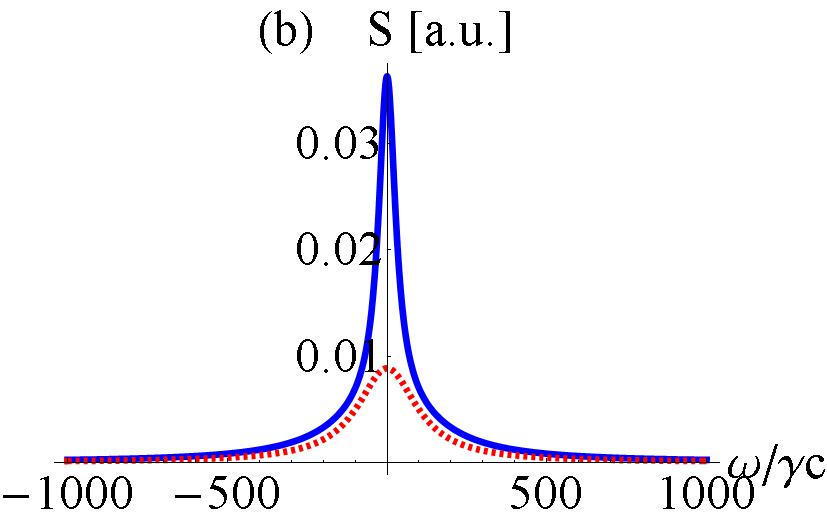}
\caption{\small{(color online). Atom fluorescence spectrum, Eq. (\ref{corr}), for modulation parameters $z=3,m=2$, and with $\omega_a=10\kappa$, $n_a=10^3$. (a) the spectrum under the influence of modulation (blue solid line) is compared to that in the absence of modulation (red dashed line, multiplied here by factor 5). The modulated case has a much narrower spectrum. (b) the spectrum in the modulated case (blue solid line) compared with a Lorentzian of width $\gamma(N+0.5)$ (red dashed line). It is apparent that the shape of the spectrum in the modulated case is not that of a single Lorentzian, as indeed can be seen in Eq. (\ref{corr}).
 }} \label{fig3}
\end{center}
\end{figure}

\subsection{Resonance fluorescence}
The resonance fluorescence spectrum of an atom in a squeezed reservoir was treated in \cite{CAR} and extended to the case of a thermal squeezed reservoir in \cite{VIET}. The atom is driven by a resonant strong (classical) field with Rabi frequency $\Omega e^{i \theta}$, and the resulting fluorescence spectrum is given by \cite{DF,VIET},
\begin{eqnarray}
S(\omega)&\propto& \frac{\gamma_{\phi}}{(\omega-\omega_a)^2+\gamma_{\phi}{^2}}
\nonumber\\
&&+\left[\frac{\Gamma_{\phi}}{(\omega-\omega_a+\Omega)^2
+\Gamma_{\phi}{^2}}+\frac{\Gamma_{\phi}}{(\omega-\omega_a-\Omega)^2+\Gamma_{\phi}{^2}}\right],
\nonumber\\
\gamma_{\phi}&=&\gamma\left(N+|M|\cos\phi+\frac{1}{2}\right),
\nonumber\\
\Gamma_{\phi}&=&\frac{3}{2}\gamma\left(N-\frac{1}{3}|M|\cos\phi+\frac{1}{2}\right),
\label{S}
\end{eqnarray}
with $\phi=2(\theta-\varphi)$. The spectrum is comprised of the three well known Lorentzian peaks centered at $\omega_a$ and $\omega_a \pm \Omega$ \cite{MOL}, but here it is modified by squeezing. In fact, we obtain a phase-dependent phenomenon, namely the widths of the fluorescence peaks, $\gamma_{\phi}$ and $\Gamma_{\phi}$, change when the phase of the strong field is varied. The phase dependence is only due to squeezing and provides a very good evidence of the effective squeezing induced by the modulation. In Fig. 4 we plot the central peak of the spectrum, a Lorentzian of width $\gamma_{\phi}$, for $z=1,3$ and different values of $\phi$. Taking the case $z=1$ for example, and beginning with $\phi=\pi/2$, $|M|\cos\phi=0$ and the only effect of the modulation is the modification of the reservoir modes occupancy $N$ with respect to the thermal one $n_a$. This results in a narrower peak, $\gamma_{\phi}=\gamma(N+0.5)=208\gamma_c$ instead of $1000.5\gamma_c$ without modulation. When $\phi$ is varied to $0$, $\gamma_{\phi}$ gets a positive contribution from the squeezing term $|M|\cos\phi=|M|$ and as a result the peak broadens to $\gamma_{\phi}=\gamma_y=309\gamma_c$. The narrowest peak is obtained for $\phi=\pi$ for which $\gamma_{\phi}=\gamma_x=106\gamma_c$.
\begin{figure}
\begin{center}
\includegraphics[scale=0.21]{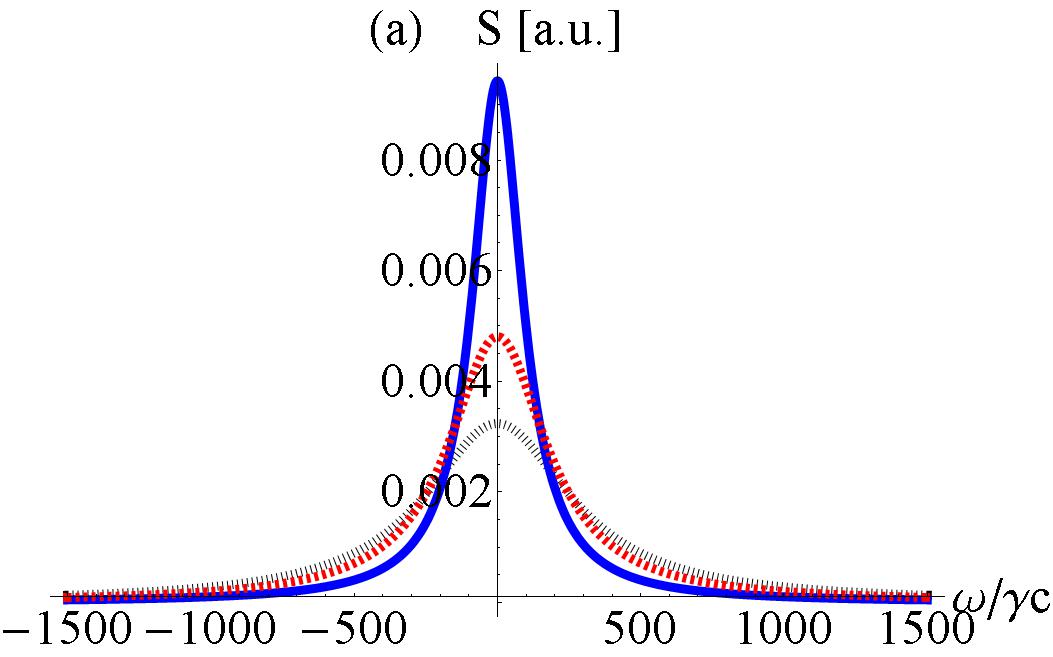}
\includegraphics[scale=0.21]{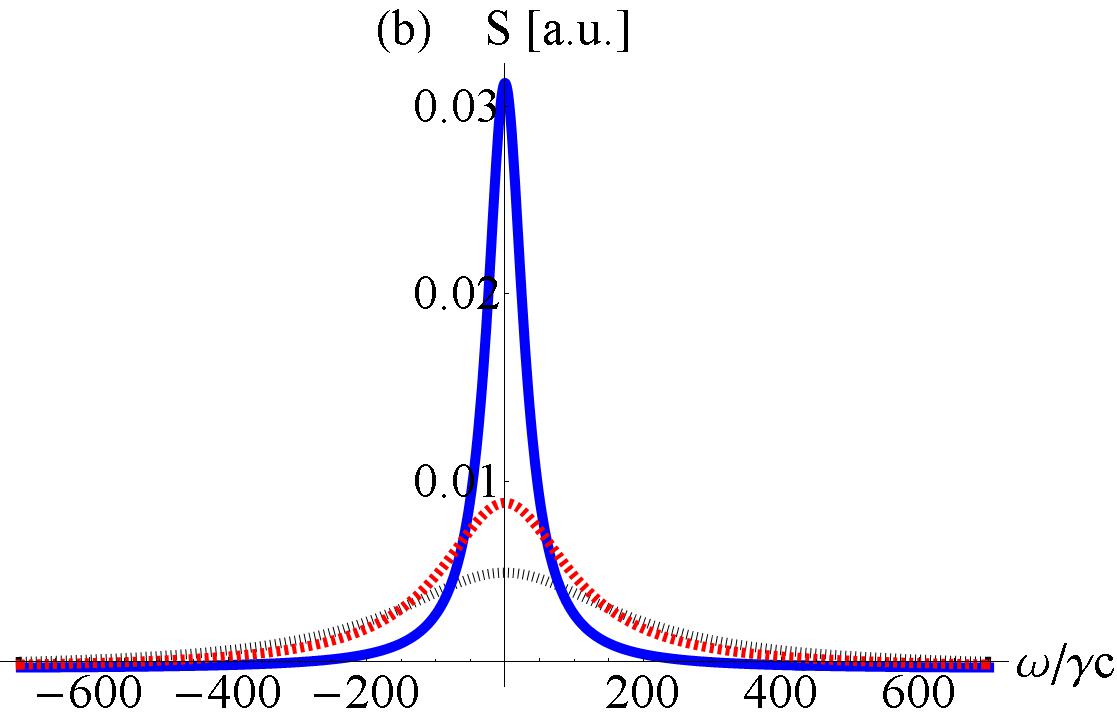}
\caption{\small{(color online). Resonance fluorescence spectrum under the influence of modulation and effective squeezing, as a function of the relative driving field phase $\phi$, for $\phi=\pi$ (blue solid line), $\phi=\pi/2$ (red dashed line) and $\phi=0$ (black dotted line). The modulation parameters are: (a) $z=1,m=2$, and (b) $z=3,m=2$. As in Figs. 2 and 3, $\omega_a=10\kappa$ and $n_a=10^3$ are taken.
 }} \label{fig4}
\end{center}
\end{figure}

\section{Experimental considerations}
We need to find a dipole transition, whose resonant energy $\omega_a$ can be controlled such that the two levels are still well separated from others. Moreover, in the sinusoidal modulation considered in Sec. V, the amplitude and frequency must exceed $\omega_a$. This leads us to consider two almost degenerate states whose energy splitting may be controlled by oscillating fields in the microwave regime. Since the effects discussed here involve coupling to radiation, it is also preferable that the dipole moment of the transition be large, as for Rydberg levels.

Let us now discuss the timescale of the above phenomena. Both the dipole dynamics and the fluorescence spectral widths are appreciable for $t>\gamma_x^{-1},\gamma_y^{-1}$. The two rates $\gamma_{x,y}$ become distinct when there is squeezing, and in the cases discussed here, are of the order of $0.1\gamma_T$ to $\gamma_T$, with $\gamma_T \equiv \gamma_c n_a \approx \gamma_c T/(\hbar\omega_a)$, where $T$ is the photon-reservoir temperature in energy units, and $\gamma_T$ is the corresponding width without modulation. Hence, we wish to observe phenomena at times $t >\gamma_T^{-1}$.

Recalling the atomic damping in the cavity, $\gamma_c$ from Eq. (\ref{GTa}), we have $\gamma_T=2\pi\frac{4g^2}{\kappa}\frac{T}{\hbar\omega_a}$, with $g=d\sqrt{\frac{\omega_a}{2\epsilon_0\hbar V}}$. Here $d$ is the dipole matrix element of the TLA transition and $V\sim\lambda_a^3$ is the cavity mode volume, taken here to be of the order of the transition wavelength cubed. We thus obtain
\begin{equation}
\gamma_T\approx\frac{3}{2\pi}\gamma_0\frac{T}{\hbar\kappa},
\label{gT}
\end{equation}
where $\gamma_0=\frac{d^2\omega_a^3}{3\pi\epsilon_0\hbar c^3}$ is the atomic transition spontaneous emission rate in free space. The above expression shows that the considered effects grow with the temperature divided by the cavity linewidth, so that high temperatures and narrow-linewidth cavities (still in the bad-cavity limit) are preferable.

Finally, let us recall the conditions of validity for some of the approximations taken in our analysis. The derivation of the reservoir spectrum in the bad-cavity limit requires $\omega_a\gg\kappa$ and $\kappa\gg g,\gamma_0$. The long-time limit in Sec. IV B is valid for $t\gg \kappa^{-1}$, i.e. in our case we demand $\gamma_T\ll\kappa$.

\emph{Example: Rydberg levels}.---
Consider the circular Rydberg level $|e\rangle=|n=51,l=50,m=50\rangle$. By dipole selection rules, it can only decay to the lower circular level $|s\rangle=|n=50,l=49,m=49\rangle$ or to the almost degenerate level $|g\rangle=|n=51,l=49,m=49\rangle$. We choose $|e\rangle$ and $|g\rangle$ as our excited and ground states, respectively, and by a static magnetic field can induce an energy (frequency) shift, of say, $1$GHz (the $s\leftrightarrow e$ transition frequency is $51.1$GHz). Another magnetic field, this time RF, is used for the modulation. It oscillates at a frequency of $2$GHz and is strong enough to induce an energy shift of a few GHz. For example, a scheme for atom fluorescence may consist of first preparing the atom in state $|e\rangle$, then applying modulations and measuring the fluorescence spectrum. Since the atom may also decay to the lower circular state $|s\rangle$, the spectrum may have an additional peak at $51$GHz, but the interesting peak for our purposes is the one at $\omega_a/(2\pi)=1$GHz.

To evaluate the typical timescale of such an experiment, we calculated the dipole matrix element of the above $g\leftrightarrow e$ transition by assuming that for such high $n$ numbers the electronic wavefunctions are hydrogenic (see Appendix D). This is such, since in these large $n$'s the last electron is far from the rest of the atom and effectively "sees" a unit positive charge like the Hydrogen nucleus. We obtained, e.g. for the $x$ component of the dipole, $d=382e a_0$, where $e$ is the electron charge and $a_0$ the Bohr radius. The spontaneous emission rate of the transition then becomes $\gamma_0\sim1.1\times 10^{-5}$Hz, and by taking $\kappa=2\times10^5$Hz, we obtain, at room temperature $T=300$K, $\gamma_T\approx2162$Hz. This means that the experiment should last $0.4$ms to $4$ms. We also verified that the conditions of validity mentioned above are satisfied. The effect can be further enhanced by effectively increasing the reservoir temperature by shining on the atom hot thermal radiation, or the radiation from any other strong incoherent source, which is wideband around $\omega_a$.

\section{Conclusions}
 This study used the apparent equivalence between counter-rotating terms in the reservoir-atom interaction and the relaxation of an atom coupled to a squeezed reservoir, to show how to engineer a thermal-reservoir so that it emulates a squeezed-reservoir. The key point that allows us to achieve this goal is the breaking of the rotating-wave approximation even at long times, by fast and strong system energy-modulation. To this end, we derived a general master equation, without invoking the Markov and rotating-wave approximations, so as to render the effect of modulation in clear form. The master equation assumes a two-level system, yet, it is easily generalized to any system-reservoir interaction of the form $\hat{X} \hat{F}$, where $\hat{X}$ is a system operator with matrix elements between different energy states of the system and $\hat{F}$ is a reservoir operator with a continuous spectrum: e.g., for a system of a harmonic oscillator with lowering operator $\hat{a}$, $\hat{X}=\hat{a}+\hat{a}^{\dag}$, and in the master equation $\hat{\sigma_-},\hat{\sigma_+}$ are replaced by $\hat{a},\hat{a}^{\dag}$ (with a modification for the energy correction term $\Delta$). Extension of the treatment to energy-level modulation in multilevel or multipartite systems described by large angular momenta \cite{GER4} is expected to yield qualitatively new features.

 In order to illustrate our scheme for squeezed reservoir engineering, we considered the case of an atom inside a cavity, namely a Lorentzian reservoir,  in the Markov regime, and applied sinusoidal modulations. We discussed possible measurable effects in atomic-dipole dephasing and fluorescence, along with experimental considerations and a possible realization of a system comprised of two atomic Rydberg levels.

\acknowledgements
We would like to thank  Ido Almog and Serge Rosenblum for fruitful discussions. We acknowledge the support of DIP, ISF and the Wolfgang Pauli Institute (E.S.).

\appendix
\section{Emulating a quantum squeezed reservoir}
In Sec. II, we found the resulting effective reservoir parameters $M$ and $N$ for case \emph{c}, where modulations are considered and only the $\varepsilon_0$ and $\varepsilon_{-2}$ terms exist. Looking at their difference,
\begin{equation}
N-|M|\propto(|\varepsilon_0|-|\varepsilon_{-2}|)\left[n(|\varepsilon_0|-|\varepsilon_{-2}|)-|\varepsilon_{-2}|\right],
\end{equation}
we can see that quantum squeezing, i.e. $|M|$ that satisfies $N<|M|<\sqrt{N(N+1)}$, becomes largest for $n=0$, i.e. for zero temperature, when the original reservoir has no thermal excitations. This is indeed what we observe for the Lorentzian reservoir in Sec. IV and the modulation from Sec. V. Taking $n_a=0$, $\omega_a=10\kappa$ and modulation parameters $m=2$ and $z=1$, nearly perfect quantum squeezing of $M=0.983\sqrt{N(N+1)}$ is achieved, with $\gamma_x=0.053\gamma_c$ and $\gamma_y=0.154\gamma_c$. The problem remains to find a realization using present-day technology. As mentioned before, since fast and strong modulation is more suitable for microwave frequencies, and since here we saw that sufficiently low temperatures are needed for $n_a\ll 1$, it is natural to consider superconducting qubits in circuit QED cavities \cite{BLA}. Then for $\omega_a=2\times10^5\kappa=2\pi\times2$GHz and typical dipole matrix elements $d=10^4ea_0$ we find $\gamma_c\approx5.7$KHz. The duration of the experiment is set by $\gamma_x^{-1}\approx3.28$ms, which is incompatible with the superconducting qubit dephasing time which currently ranges between $1\mu$s and $10\mu$s. Future technology may allow emulating a quantum squeezed reservoir by the method described in this paper. However at present this appears to be very challenging.

\section{Note on the derivation of the master equation}
Here we would like to show how to get equation (\ref{B3b}) from Eq. (\ref{B3a}). Beginning with the time integration over $t'$ and denoting $\nu\equiv-\omega_a-\nu_q$, we change variables to $\tau=t-t'$ and get
\begin{equation}
\int_0^t dt' e^{-i(\omega-\nu)(t-t')}=\int_{-t}^t d\tau \Theta(\tau) e^{-i(\omega-\nu)\tau},
\label{A1}
\end{equation}
with $\Theta(x)$ the Heaviside step function. By contour integration one can obtain the relation,
\begin{equation}
\Theta(\tau) e^{-i(\omega-\nu)\tau}=-\lim_{\eta\rightarrow 0^+} \frac{1}{2\pi i}\int_{-\infty}^{\infty}d\omega'\frac{e^{-i\omega' \tau}}{\omega'+i\eta-(\omega-\nu)}.
\label{A2}
\end{equation}
Inserting Eqs. (\ref{A1},\ref{A2}) into the double integral in Eq. (\ref{B3a}) (denoted here $I$), we get
\begin{equation}
I=\int_{-\infty}^{\infty}d\omega' \int_{-t}^t d\tau \frac{e^{-i\omega' \tau}}{2\pi i} \lim_{\eta\rightarrow 0^+}\int_{-\infty}^{\infty}d\omega \frac{-G_n(\omega)}{\omega'+i\eta-(\omega-\nu)},
\label{A3}
\end{equation}
with $G_n(\omega)=G_0(\omega)[n(\omega)+1]$. Using the relation
\begin{equation}
\lim_{\eta\rightarrow 0^+}\frac{-1}{\omega'+i\eta-(\omega-\nu)}=i\pi\delta(\omega-(\omega'+\nu))+P\frac{1}{\omega-(\omega'+\nu)},
\label{A4}
\end{equation}
where $P$ denotes the principal value under integration, and defining the sinc function $2\pi \delta_t(\omega')=\int_{-t}^t d \tau e^{-i \omega' \tau}$, we finally obtain,
\begin{equation}
I=\int_{-\infty}^{\infty}d\omega'\delta_t(\omega')\left[iP\int_{-\infty}^{\infty}d\omega \frac{G_n(\omega)}{\omega+(\omega'-\nu)}-\pi G_n(\omega'+\nu)\right].
\label{A5}
\end{equation}
Noting Eq. (\ref{B3c}), this is just the expression in Eq. (\ref{B3b}) with $\omega$ replaced by $\omega'$.

\section{The spectrum of a cavity reservoir}
Our aim here is to derive the correlation functions in Eq. (\ref{a}), for a cavity-mode in the bad-cavity limit. We begin with the model Hamiltonian $H=H_S+H_R+H_{SR}$ where the system ($H_S$) is the atom and cavity mode ($\hat{a}$), the reservoir ($H_R$) is the electromagnetic modes that the cavity is leaking to ($\hat{b}_j$) and the perpendicular modes that the atom decay to ($\hat{r}_l$), and their interaction is described by $H_{SR}$,
\begin{eqnarray}
H_S&=&\hbar \omega_a \hat{a}^{\dag}\hat{a}+0.5\hbar\hat{\sigma}_z\omega_a+\hbar g (\hat{a}+\hat{a}^{\dag})(\hat{\sigma}_++\hat{\sigma}_-),
\nonumber\\
H_R&=&\sum_j\hbar\omega_j\hat{b}_j^{\dag}\hat{b}_j+\sum_l\hbar\omega_l\hat{r}_l^{\dag}\hat{r}_l,
\nonumber\\
H_{SR}&=&\hbar\sum_j \eta_j(\hat{a}+\hat{a}^{\dag})(\hat{b}_j+\hat{b}_j^{\dag})
\nonumber\\
&&+\hbar\sum_l g_l(\hat{\sigma}_-+\hat{\sigma}_+)(\hat{r}_l+\hat{r}_l^{\dag}).
\label{B1}
\end{eqnarray}
By writing the Heisenberg equations for $\hat{a}$ and $\hat{b}_j$ and inserting the later into the former we obtain an equation for the envelope of the cavity mode $\tilde{a}(t)=\hat{a}(t)e^{i\omega_a t}$,
\begin{eqnarray}
&&\dot{\tilde{a}}=-\sum_j|\eta_j|^2\int_0^t dt'\tilde{a}(t')e^{-i(\omega_j-\omega_a)(t-t')}+\hat{F}(t)-ig\hat{\mu},
\nonumber\\
&&\hat{F}(t)=-i\sum_j\eta_j\hat{b}_j(0)e^{-i(\omega_j-\omega_a)t},
\label{B2}
\end{eqnarray}
where $\hat{\mu}=\hat{\sigma}_++\hat{\sigma}_-$ and $\hat{F}$ is the Langevin force induced on the cavity mode by the outside modes. As in standard Heisenberg-Langevin theory \cite{SCU}, we now take the Markov approximation, i.e. we assume that the typical timescale of the cavity mode dynamics, dictated by $\kappa\equiv D_b(\omega_a)|\eta(\omega_a)|^2$ with $D_b(\omega)$ the density of outside modes, is much longer than the inverse of the bandwidth of $D_b(\omega)|\eta(\omega)|^2$. Typically, for a power-law density of states, we may take this bandwidth to be $\omega_a$, so in fact we assume here $\kappa\ll\omega_a$. Then, $\tilde{a}$ can be taken out of the integral and by averaging, assuming $\langle\hat{b}_j(0) \rangle=0$, we get,
\begin{equation}
\langle\dot{\tilde{a}}\rangle=-\frac{\kappa}{2}\langle\tilde{a}\rangle-ig\langle\hat{\mu}\rangle.
\label{B3}
\end{equation}
The typical timescale for changes in the atomic operators is dictated by the decay rate to the perpendicular modes, which is similar to that in free-space, $\gamma_0$, and by $g$. By assuming the bad-cavity limit, i.e. $\kappa\gg g,\gamma_0$, and noting that $\langle\hat{\mu}\rangle\leq O(1)$, we have $\langle\dot{\tilde{a}}\rangle\approx-\frac{\kappa}{2}\langle\tilde{a}\rangle$, and thus obtain
\begin{equation}
\langle\tilde{a}\rangle(t)\approx \langle\tilde{a}\rangle(0) e^{-\frac{\kappa}{2}t}.
\label{B4}
\end{equation}
Finally, by the quantum regression theorem \cite{CARb} we have
\begin{eqnarray}
\langle \tilde{a}(t)\tilde{a}^{\dag}(t+\tau)\rangle&=&\langle \tilde{a}(t)\tilde{a}^{\dag}(t)\rangle e^{-\frac{\kappa}{2} |\tau|}=(n_a+1)e^{-\frac{\kappa}{2} |\tau|}
\nonumber\\
\langle \tilde{a}^{\dag}(t)\tilde{a}(t+\tau)\rangle&=&\langle \tilde{a}^{\dag}(t)\tilde{a}(t)\rangle  e^{-\frac{\kappa}{2} |\tau|}=n_a e^{-\frac{\kappa}{2} |\tau|},
\label{B5}
\end{eqnarray}
where in the last step we assumed stationarity of the correlations, and took the thermal state as the stationary state, with $n_a$ the Planck distribution for $\omega_a$.

\section{Dipole matrix element calculation}
We would like to calculate the dipole matrix element $d$ between the states $|e\rangle=|n,n-1,n-1\rangle$ and $|g\rangle=|n,n-2,n-2\rangle$, with $n=51$. Let us first derive an expression for a general $n$. Clearly the $z$ component of the dipole vanishes since $\Delta m=1$, so let us calculate $d_x=e r \sin \theta \cos \phi$, with the usual convention of spherical coordinates. In terms of the Hydrogen wavefunctions $R_{nl}(r)Y_{lm}(\Omega)$, where $\Omega$ is the solid angle, the matrix element is written,
\begin{eqnarray}
&&\frac{d}{e}=RA
\nonumber\\
&&R=\int_0^{\infty}dr r^3 R^{\ast}_{n,n-1}(r)R_{n,n-2}(r)
\nonumber\\
&&A=\int d\Omega Y^{\ast}_{n-1,n-1}(\Omega)\sin\theta\cos\phi Y_{n-2,n-2}(\Omega).
\label{C1}
\end{eqnarray}
We begin with the radial part $R$. Recalling the Hydrogen radial functions \cite{GRI}, we find
\begin{eqnarray}
R_{n,n-1}(r)&=&\sqrt{\left(\frac{2}{n a_0}\right)^3\frac{1}{2n(2n-1)!}}e^{-r}{na_0}
\nonumber\\
&&\times\left[-\left(\frac{2r}{na_0}\right)^{n-1}+(2n-2)\left(\frac{2r}{na_0}\right)^{n-2}\right]
\nonumber\\
R_{n,n-2}(r)&=&\sqrt{\left(\frac{2}{n a_0}\right)^3\frac{1}{2n(2n-2)!}}e^{-r}{na_0}\left(\frac{2r}{na_0}\right)^{n-1}.
\nonumber\\
\label{C2}
\end{eqnarray}
Performing the integrations in $R$, we obtain
\begin{eqnarray}
R&=&\sqrt{\frac{1}{(2n-1)!(2n-2)!}}\frac{1}{2}
\nonumber\\
&&\times\left[(n-1)\Gamma(2n+1)-\frac{1}{2}\Gamma(2n+2)\right]a_0,
\label{C3}
\end{eqnarray}
where $\Gamma(x)$ is the Gamma function. For $n=51$ we get $R=-768.815a_0$. The angular part $A$ is calculated by first noting that $\sin\theta\cos\phi=\sqrt{\frac{2\pi}{3}}(Y_{1,-1}-Y_{1,1})$ and then using
\begin{equation}
\int d\Omega Y^{\ast}_{lm}Y_{1,q}Y_{l'm'}=\sqrt{\frac{3(2l'+1)}{4\pi(2l+1)}}C_{l,0;l',1,0,0}C_{l,m;l',1,m',q},
\label{C4}
\end{equation}
where $C_{J,M;j_1,j_2,m_1,m_2}$ are the Clebsch-Gordan coefficients. We thus obtain,
\begin{eqnarray}
A&=&\sqrt{\frac{2n-3}{2(2n-1)}}C_{n-1,0;n-2,1,0,0}
\nonumber\\
&&\times \left[C_{n-1,n-1;n-2,1,n-2,-1}-C_{n-1,n-1;n-2,1,n-2,1}\right].
\nonumber\\
\label{C5}
\end{eqnarray}
For $n=51$ we get $A=0.497$, so finally we find $d=-382.101e a_0$.

\end{document}